\def\preprint{0}		
\def\preprint{1}		

\if\preprint1
	\documentclass{article}
	\usepackage{mn2e}
	\usepackage{astrop-bib}
	\usepackage{times}
	\usepackage{graphicx}
	\usepackage{calc}

\else
	\documentclass{article}
	\usepackage[referee]{mn2e}
	\usepackage{astrop-bib}
	\usepackage{times}
	\newcommand{\includegraphics}[2][]{\relax} 
\fi

\newcommand{\bCen}{\mbox{$\beta$\,Cen}}
\newcommand{\half}{{\textstyle\frac{1}{2}}}
\newcommand{\comment}[1]{\relax}

\ifnfsstwo

\fi

\ifnfssone
  \newmathalphabet{\mathit}
    \addtoversion{normal}{\mathit}{cmr}{m}{it}
    \addtoversion{bold}{\mathit}{cmr}{bx}{it}

\fi

\ifoldfss

\fi

\loadboldmathitalic
\loadboldgreek

\title[\bCen: a binary $\beta$\,Cephei star]
{Interferometry and spectroscopy of $\bbeta$\,Cen: a  $\bbeta$\,Cephei
star in a binary system\thanks{Based on observations obtained at the
Anglo-Australian Observatory and the European Southern Observatory, Chile}}
\author[J.~G.~Robertson et al.]
       {J.~G.~Robertson,$^1$\thanks{E-mail: \tt jgr@physics.usyd.edu.au}
	T.~R.~Bedding,$^1$
	C. Aerts$^{2}$\thanks{Postdoctoral Fellow, Fund for Scientific
		Research, Flanders (Belgium)},
	C. Waelkens$^{2}$,
	R.~G.~Marson$^{1,3}$ and\cr
	J.~R.~Barton$^4$\\
	$^1$School of Physics, University of Sydney 2006, Australia\\
	$^2$Instituut voor Sterrenkunde, Celestijnenlaan 200\,B, B-3001
	Leuven, Belgium\\
	$^3$Current address: NRAO Array Operations Center, P.O. Box 0,
	Socorro NM 87801, USA\\
	$^4$Anglo-Australian Observatory, P.O. Box 296, Epping 2121,
	Australia\\
}
\pubyear{1998}

\begin{document}

\maketitle

\begin{abstract}
\bCen{} is a bright $\beta$\,Cephei variable and has long been suspected
to be a binary.  Here we report interferometric observations with the
3.9-m Anglo-Australian Telescope at a single epoch which show the star to
be a binary with components  separated by 15 milliarcsec and having
approximately equal luminosities at 486 nm.  We also present
high-resolution spectra taken over five nights with the ESO CAT which show
\bCen{} to be a double-lined spectroscopic binary.  We identify two
pulsation frequencies in the primary.  Further spectroscopic and
interferometric studies of this double star should allow determination of
its orbital parameters and component masses.
\end{abstract}


\section{Introduction}

\bCen{} (HR 5267) is a bright early-type southern star.  It has been of
interest to observers as a $\beta$\,Cephei variable, as a probe of the
intervening interstellar medium, and also as a stellar X-ray source.
\bCen{} was observed by the Narrabri Stellar Intensity Interferometer in
1963 as a target for commissioning observations of that instrument
\cite{Han}.  The maximum correlation observed was, however, only half the
expected value. This could be explained if \bCen{} were a double star with
components of similar brightness.  The Intensity Interferometer results
could not directly demonstrate the binarity of \bCen{}, nor determine the
separation, because the angular separation was resolved by  the individual
6.5 m reflectors. But a value greater than about a few hundredths of an
arcsecond would be expected. The alternative explanation of a single star
surrounded by a halo was considered unlikely since no narrow emission or
absorption lines were seen in the spectrum \cite{S+R68}.

Binarity was also suspected by \citeone{Bre67} and \citeone{S+R68} from
radial velocity measurements.  The latter authors were seeking to clarify
the low correlation found by the Intensity Interferometer.  They found
velocity variations with a short period (0.135\,d), indicative of a
$\beta$\,Cephei variable, and a long period (352\,d), ascribed to orbital
motion in a binary system.  A more comprehensive spectroscopic analysis was
made by \citeone{Lom75}, who classed the star as a single-lined
spectroscopic binary and gave the short period as 0.157\,d.

Here we report further observations of \bCen, both interferometric and
spectroscopic, which conclusively demonstrate binarity.  We have also
identified two pulsation frequencies in the primary from radial-velocity
variability.  Further observations will be needed to find the parameters of
the orbit.

\if\preprint1
	\begin{table*}
\begin{flushleft}
\caption[]{\label{table.obs} Journal of MAPPIT observations of \bCen}
\begin{tabular}{llllllll}

Date & $\lambda$/$\Delta\lambda$ & Mask &  Longest & Hole & 
		Detector& Observed position angles & Seeing\\
     &           & pattern & baseline$^*$ & size$^*$ & field & (North through East)& \\
(UT) & (nm)& & (m)& (cm) & (arcsec) & (degrees)& (arcsec) \\ \hline
1995/1/12 & 650/40 &~A& 3.28 & 4.7 & 3.3 x 5.8 & 87.8, $-$31.1, 29.8 & 1.1 \\
1995/1/12 & 650/40 &~B& 3.67 & 4.7 & 3.3 x 5.8 & 5.1, 60.2, 11.1, $-$57.8 & 1.2 \\
1995/1/14 & 486/10 &~B& 3.67 & 7.9 & 2.3 x 4.1 &
	 $-$1.1, $-$90.0, $-$89.4, 45.9, $-$42.8, 4.2, $-$60.3, 
				& $\sim$1 (cloud-affected)\\
          & & & & & & 74.8, $-$13.5, 33.6, 82.2, $-$3.0, $-$71.9, $-$71.3 & \\
\hline
\end{tabular}\\
$^*$As projected on primary mirror 
\end{flushleft}
\end{table*}

\fi

\section{Interferometric Measurements}

\subsection{Observations}

We have carried out observations of \bCen{} using aperture masking, which
involves modifying the telescope pupil with a mask \cite{HMT87}.  Our
aperture masks contain a small number of non-redundantly spaced holes, each
of size comparable to or smaller than the Fried length $r_0$. Compared with
a fully-filled aperture, as used in speckle interferometry, such
observations provide better calibrated measurements and reach the full
diffraction limit. This comes at the expense of a brighter limiting
magnitude and less complete spatial-frequency coverage, neither of which is
important for a bright double star.  As we show in this paper, the
Non-Redundant Masking (NRM) method gives very clear resolution of a binary
at 15\,mas from observations with a 4 m-class telescope, whereas speckle
interferometry with similar telescopes is limited to separations of greater
than 30\,mas
\cite{McA97}.

We observed \bCen{} in 1995 January with the MAPPIT interferometer (Masked
APerture-Plane Interference Telescope; \citebare{BRM94b}), which is
situated at the coud\'e focus of the 3.9-m Anglo-Australian Telescope
(AAT).  The instrumental setup was the same as that described by
\citeone{BZvdL97}.  Table~\ref{table.obs} gives particulars of the
observations.  The unresolved star $\alpha^1$~Cen was observed with each of
the three wavelength/mask combinations tabulated, to provide calibration of the
fringe visibility loss due to residual atmospheric effects.  A Dove prism
was used as a field rotator to obtain a range of observed position angles,
and an atmospheric dispersion corrector was used during all observations.
The binary stars ADS~7846, ADS~7982, ADS~8573 and $\gamma$~Vir were
observed to provide a calibration of the spatial scale and position angle.

\if\preprint1
	\begin{figure}
\includegraphics[width=\the\hsize]{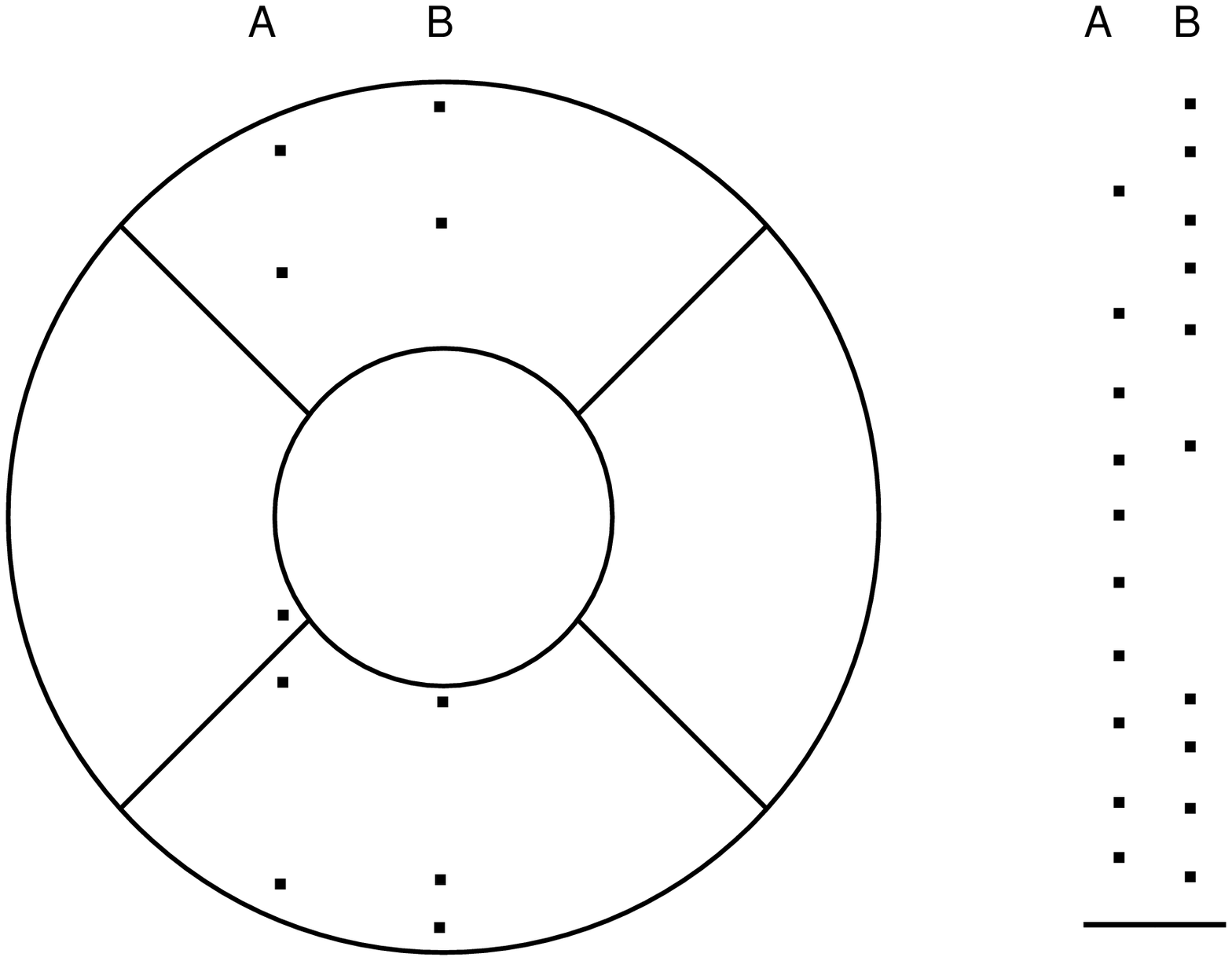}

\caption[]{\label{fig.mask} Non-redundant mask designs for MAPPIT, with the
pupil of the AAT drawn to the same scale.  The five holes in array~A have
relative positions (scaled to the longest baseline) of 0, 0.275, 0.367,
0.833 and 1.000.  The holes in array~B are at relative positions 0, 0.058,
0.275, 0.858 and 1.000.  The columns on the right show the spatial
frequency coverage of the two arrays (the line at the bottom is the
origin).  The holes as shown have sizes, as projected on the primary
mirror, of 4.7\,cm.  }

\end{figure}

\fi

Two different aperture masks were used, both having 5 square holes in a
straight line and giving 10 simultaneous non-redundant baselines.  The mask
layouts are shown in Fig.~\ref{fig.mask} (only one mask was used at a
time).  Design~A gives a nearly uniform spacing of the baselines
\cite{Mar94}, but is limited to a maximum baseline of 3.28\,m by the need
to fit the holes around the central obstruction of the telescope.  The
other design has a maximum baseline of 3.67\,m, close to the largest that
can reliably be fitted on the AAT primary mirror, but accepts a non-uniform
distribution of baselines \cite{Bed92}.  For the present observations,
where the visibility as a function of baseline is expected to take the
simple form given by a binary star, both masks are suitable.  The longer
baseline mask and shorter wavelength were used in later runs to help ensure
that we observed the null of the binary visibility curve.

The detector was a Thomson CCD, with a window of 230 columns by 400 rows.
The CCD was used in time series mode with full column binning
\citeeg{BHB90}, giving a $230\times1$ readout every 13\,ms.  In this mode
it was possible to observe using a 100\% duty cycle, i.e., with the shutter
open continuously during each 130-s run.  Full column binning was possible
without smearing the fringes because the fringes were arranged to lie
accurately along columns.

\subsection{Data analysis}

The interferometric data consisted of 21 runs spaced over the full range
of position angles (see Table 1), and with 10 baselines for every run.
Power spectra were computed for each of the 10,000 exposures in a run, and
then combined to give an average power spectrum for that run. The average
power spectra exhibited 10 peaks, from the 10 non-redundant baselines.
The heights of these peaks were measured in order to find the observed
values of $V^2$ (the square of the fringe visibility).  These were
calibrated by dividing by the corresponding values from the observation of
the calibration star.  An advantage of NRM is that the peak heights can be
measured relative to their local background, thereby allowing for noise
variance contributions to the power spectra.

The individual stellar discs of hot stars  cannot be resolved by baselines
limited to 4 m. \bCen{} (spectral type B2) thus falls in the class of
binary stars with the individual stellar discs unresolved, for which $V^2$
is given by
\begin{equation}
  V^2 = \frac{1}{(1+\beta)^2}
	\left[ \beta^2 + 1 +
		2\beta\cos\left\{(2\pi\theta d \cos\psi)/\lambda\right\}
	\right], 		\label{eq.general}
\end{equation}
where $\beta$ is the brightness ratio of the two components, $\theta$ is
their angular separation, $\psi$ is the angle on the plane of the sky
between the line joining the two components and the projection of the
interferometer baseline, $\lambda$ is the observing wavelength and $d$ is
the interferometer baseline length \cite{HDA67}.

\if\preprint1
	\begin{figure}

\includegraphics[width=\the\hsize,bb=89 372 556 694]{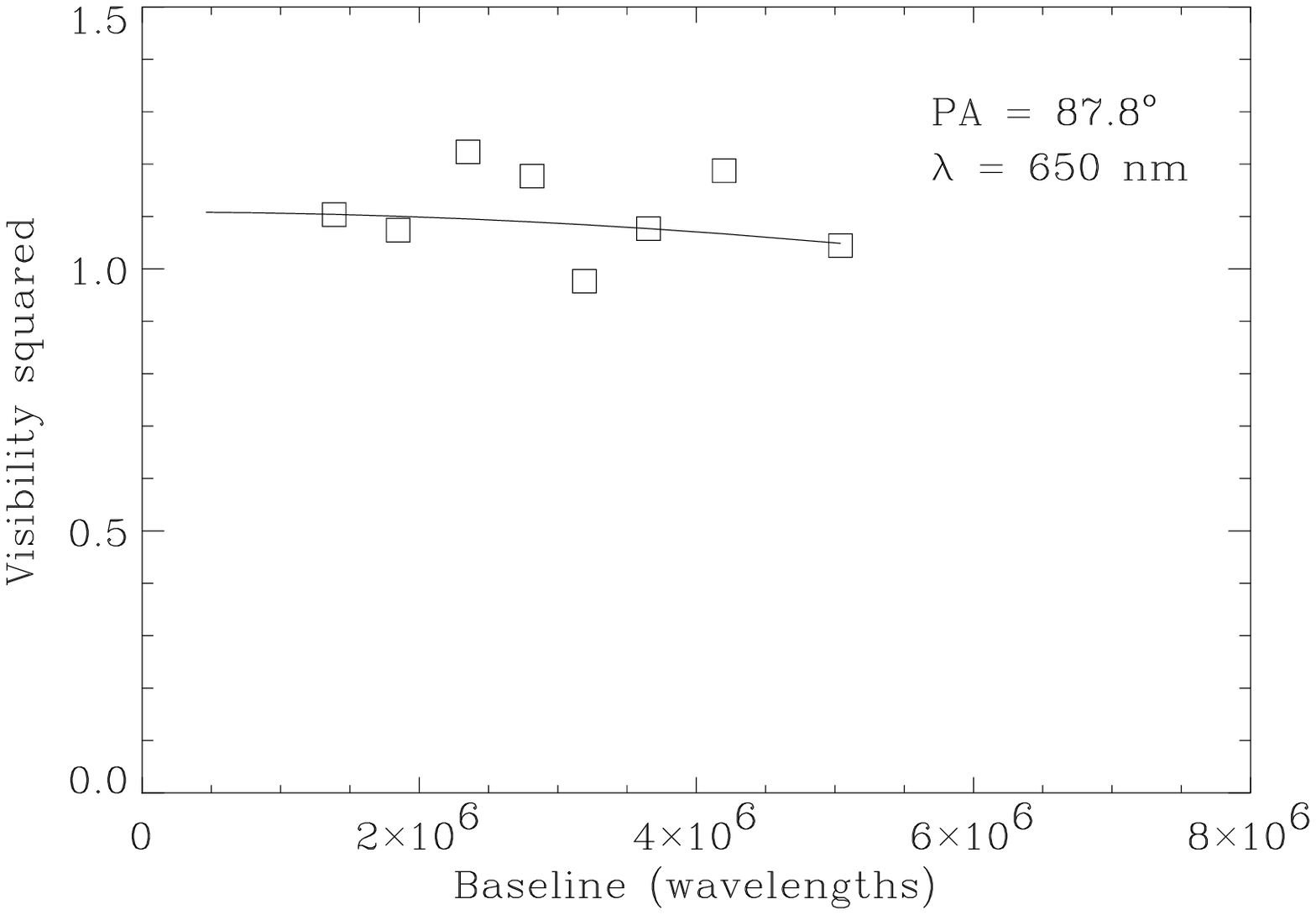}

\bigskip

\includegraphics[width=\the\hsize,bb=89 372 556 694]{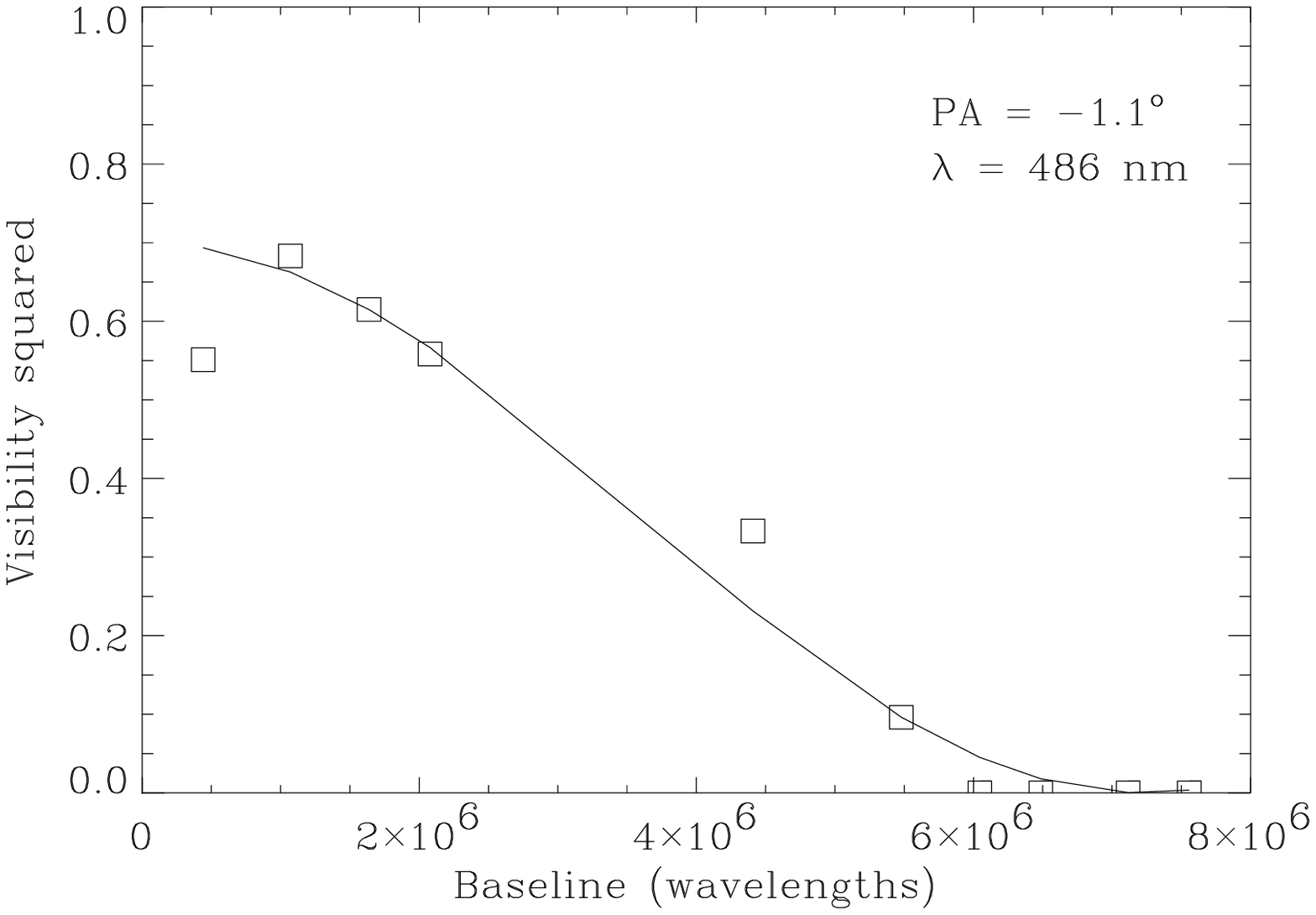}

\caption[]{\label{fig.vis} Two examples of $V^2$ fits for MAPPIT
observations of \bCen.  The symbols show measured values and the curves are
two-parameter fits, as described in the text.  }

\end{figure}

\fi

We have adopted a two-step process in fitting the functional form expected
for a binary star to the data.  The first step consisted of the separate
treatment of each calibrated run.  Figure~\ref{fig.vis} shows two examples,
with different degrees of visibility taper due to the different
angles~$\psi$.  Runs with $\psi$ near 0, i.e., those that best resolve the
double, show that $V^2$ does exhibit nulls.  This indicates that the two
stars have equal brightness to within the observational accuracy (the limit
which can be placed on the magnitude difference is discussed below).  The
model fit to each run was therefore specialised to the case of $\beta = 1$,
giving
\begin{equation}
  V^2 = \half
	\left[ 1 +
		\cos\left\{(2\pi\theta d \cos\psi)/\lambda\right\}
	\right].  		\label{eq.equal}
\end{equation}

For each run, two free parameters $A$ and $B$ were allowed, by fitting the
following model functional form to the observed values of $V^2$:
\begin{equation}
  V_{\rm mod}^2 = \half A
	\left[ 1 +
		\cos\left( B\,2\pi d/\lambda \right)
	\right].  		\label{eq.model}
\end{equation}
The parameter $B$ depends on the degree of taper of the visibilities with
baseline (large $B$ values indicate strong taper), which in turn depends on
the position angle difference,~$\psi$.  Ideally, the parameter $A$ should
be 1.0 for calibrated observations.  In practice, it was clear that the
internal consistency of the visibilities for the 10 baselines within each
run was better than the consistency between runs, and it was preferable to
allow the overall visibility normalisation of each run to float by fitting
the $A$ values.  Variations in atmospheric conditions between calibrator
and target star, including the effects of cloud on some runs, were probably
responsible.

The fitting process for each run consisted of finding the values of $A$ and
$B$ which minimised the value of the sum over the data
points of
\begin{equation}
  \left| \frac{V_{\rm obs}^2 - V_{\rm mod}^2}{V_{\rm mod}^2 + 0.2} \right|,
  \label{eq.minimize}
\end{equation}
where $V_{\rm obs}$ refers to the observed visibilities.  Minimum absolute
value rather than a least squares fit was used to obtain a more robust
treatment of the non-Gaussian errors, which arise from residual atmospheric
effects on the values of $V^2$. The reason for minimising the expression as
shown is to allow for the observed greater absolute errors at high
visibilities, giving a process intermediate between the assumptions of
constant absolute errors and constant fractional errors. With the additive
constant 0.2 in the denominator, a deviation between model and data at $V^2
\sim 1$ can be 6 times the deviation at low visibilities ($V^2 \sim 0$) for
the same contribution to the sum of misfit terms and hence equal influence
on the fitting function.

\if\preprint1
	\begin{figure}

\includegraphics[width=\the\hsize,bb=92 372 548 694]{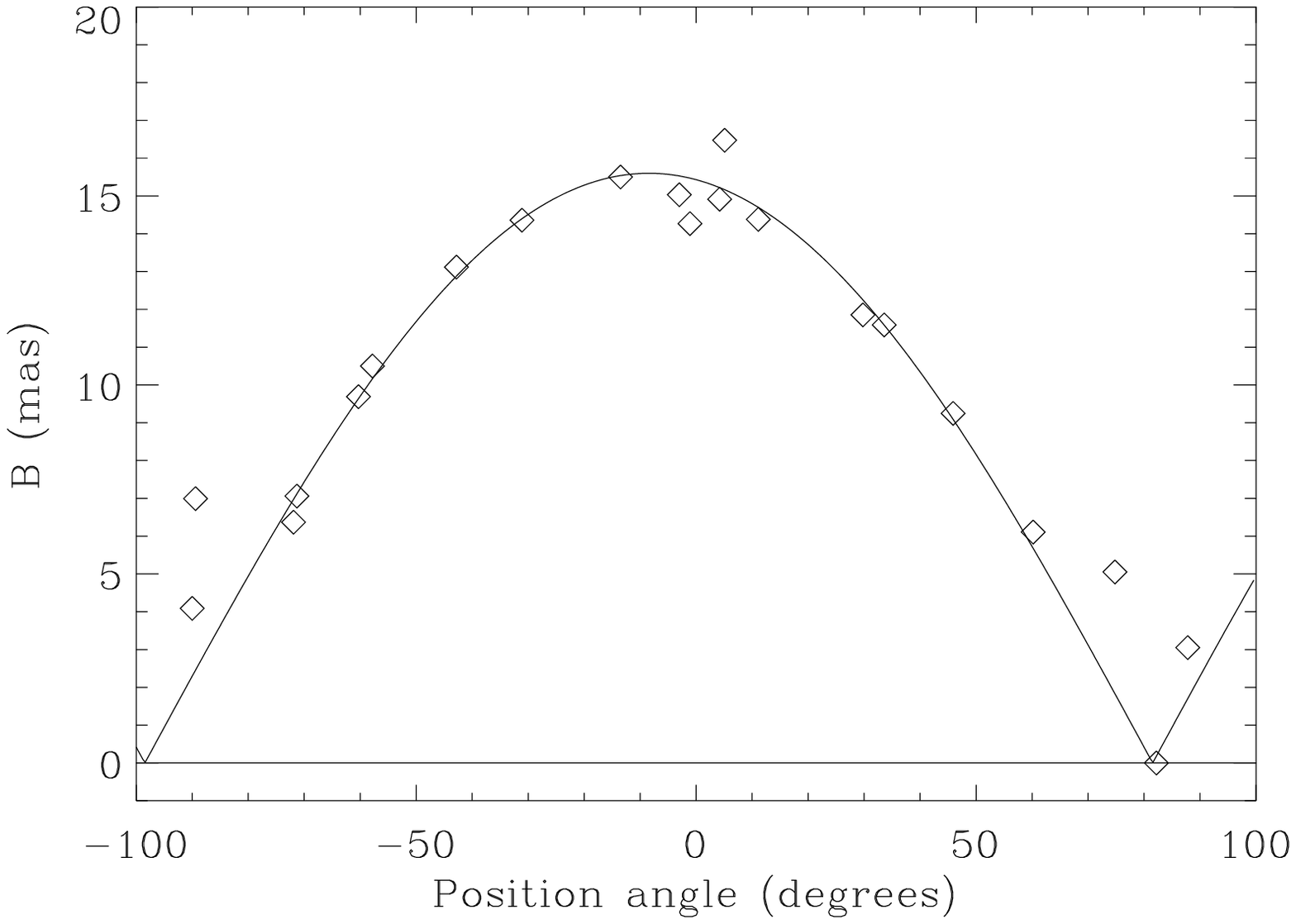}

\caption[]{\label{fig.B} Projected angular separation of \bCen{} as a
function of position angle.  Each symbol is the result of a fit to
observations at a single position angle.  The curve shows the best-fit
theoretical model for a binary star with components of equal luminosity.  }

\end{figure}

\fi

The second step in the data reduction was to make a fit to the 21 values of
the parameter $B$ as a function of the observed position angle.
Figure~\ref{fig.B} shows the result.  Breaking the fitting process into two
steps in this way enables the data quality to be clearly illustrated,
because Figure~\ref{fig.B} shows 21 independent data points, fitted by a
model which depends on only two parameters.  Note that $B$ is the angular
separation of the stars as projected along the interferometer
baseline. Consistent with Equation~(\ref{eq.equal}), the model functional
form we have fitted is
\begin{equation}
  B = \left| \theta \cos(\phi_{\rm obs} - \phi) \right|,
		\label{eq.B}
\end{equation}
where $\phi_{\rm obs}$ is the position angle on the sky at which
observations are made and $\phi$ is the position angle of the line joining
the two stars.  The parameters to be fitted are the separation $\theta$ and
the true sky position angle~$\phi$.  The observed position angles have been
folded into the range $-90^{\circ} < \phi_{\rm obs} \le 90^{\circ}$.  In
principle, closure phases could be derived from the data and used to
resolve the $180^{\circ}$ ambiguity, but this is not feasible for the present
case in which the component stars are nearly equal in brightness.

\subsection{Results}

The results in Fig.~\ref{fig.B} show clearly that the parameter $B$ follows
the expected functional form.  The points with low $B$ values have larger
uncertainties, principally because the available maximum baseline does not
then reach the visibility null or close to it.  For this reason, robust
fitting was again achieved by minimising the sum of absolute values of the
deviations between the model and data, as seen in Fig.~\ref{fig.B}.

We obtain an angular separation between the two components of $15.6 \pm
2$\,mas and a sky position angle of $-8.5^{\circ} \pm 6^{\circ}$ (with a
$180^{\circ}$ ambiguity).  The uncertainties allow for the scatter as
shown in Fig.~\ref{fig.B} (at 1$\sigma$ level), and additional systematic
errors in the angular scale and position angle calibration. Measuring
position angle from North through East, our final values for separation
and position angle are therefore $15.6 \pm 2$\,mas and either
$352^{\circ}\pm 6^{\circ}$ or $172^{\circ}\pm 6^{\circ}$.

Equation~(\ref{eq.general}) shows that the two components must have equal
brightness in order for the visibility versus baseline curve to have a
null.  Our data do show nulls where the observed position angle is
appropriate, and thus our results are consistent with two stars of exactly
equal magnitude.  To set a limit on the magnitude difference, we have
examined the regions of the nulls and find a limit of $V^2 < 0.015$
($\sim2\sigma$), which corresponds to a magnitude difference $\Delta m <
0.3$ (at 486\,nm).

Another limit to the magnitude difference can be obtained from the
Intensity Interferometer observations by \citeone{HDA74}.  The
`zero-baseline correlation' for \bCen{} was found to be $0.47 \pm 0.02$.
The minimum value predicted by theory for a binary system with unresolved
individual components is 0.5, which occurs for components of equal
brightness.  The observed value implies equal components, and with the
given uncertainty corresponds to a limit $\Delta m < 0.3$ (2$\sigma$, at
461\,nm), consistent with our result.

\section{Spectroscopic Observations}

In the course of a systematic study of line-profile variations in
$\beta\,$Cep stars started in the 1980s \cite{Aer93}, we have obtained 50
high-resolution ($R=100\,000$), high S/N ($>300$) spectra of $\beta\,$Cen
taken during five consecutive nights (1988 May 16-20).  The observations were
performed at the European Southern Observatory in Chile with the CAT/CES
instrumentation using a Reticon detector.  We observed the Si\,III line at
$\lambda 4553\,$\AA.  Integration times were typically 5 minutes, which is
less than 1\% of the pulsation period of 0.157 days (6.37\,cycles/day,
hereafter abbreviated as c/d) proposed for $\beta\,$Cen by
\citeone{Lom75}. The wavelengths and radial velocities we give are
heliocentric.

\if\preprint1
	\begin{figure}
\includegraphics[width=\the\hsize,bb=35 232 394 451]{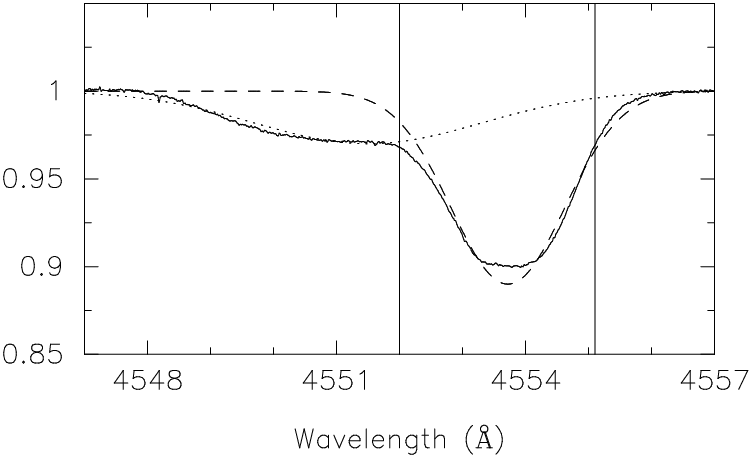}

\caption[]{\label{fig.spectrum} The average profile of the
$\lambda\,4553\,$\AA\ Si\,III line of our $\beta\,$Cen data set of 1988 May
(full line).  The dashed and dotted lines show the best Gaussian fits for
the primary and secondary respectively.  The vertical lines denote the
integration limits for calculation of the first velocity moments of
the primary's line.}

\end{figure}

\fi

\subsection{Average line profiles}

The average profile of the $4553\,$\AA\ Si\,III line is shown in
Fig.~\ref{fig.spectrum} and clearly reveals $\beta\,$Cen to be a
double-lined spectroscopic binary.  For convenience we will refer to the
star producing the stronger and narrower line, with clear $\beta$ Cep line
profile variations (see Section \ref{Line-profile variations}), as the
primary, and the other as the secondary, despite their essentially equal
magnitudes. The secondary has a broader line, presumably due to faster
rotation.

We have determined the best Gaussian fit to the average line profiles of
the primary and of the secondary component (shown as dashed and dotted
lines in Fig.~\ref{fig.spectrum}).  We derive upper limits for the
projected rotation velocities of $v\sin\,i = 120$\,km\,s$^{-1}$ and
$265$\,km\,s$^{-1}$ for the primary and secondary, respectively.  More
accurate values cannot be derived, since it is unclear from our data what
fraction of the broadening of the lines is due to pulsation.  It should be
mentioned that all radial velocity values listed in the literature so far
are based on the assumption of a single-lined spectroscopic binary and must
therefore be treated with caution.

\if\preprint1
	\begin{figure*}

 \includegraphics[angle=-90, width=10cm, bb=77 51 535 707, clip=true]
{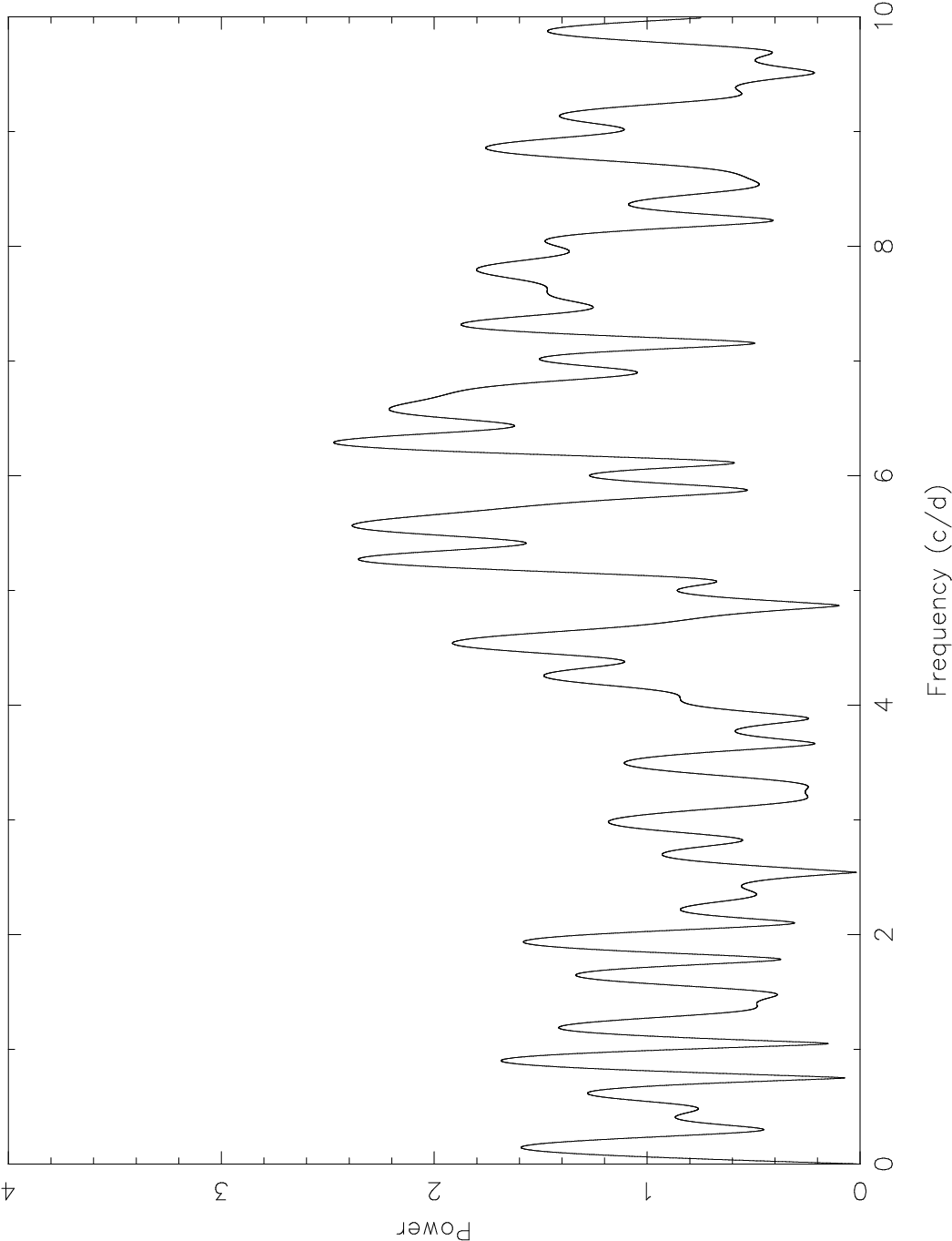}

\smallskip

\includegraphics[width=10cm,bb=23 5 482 357]
{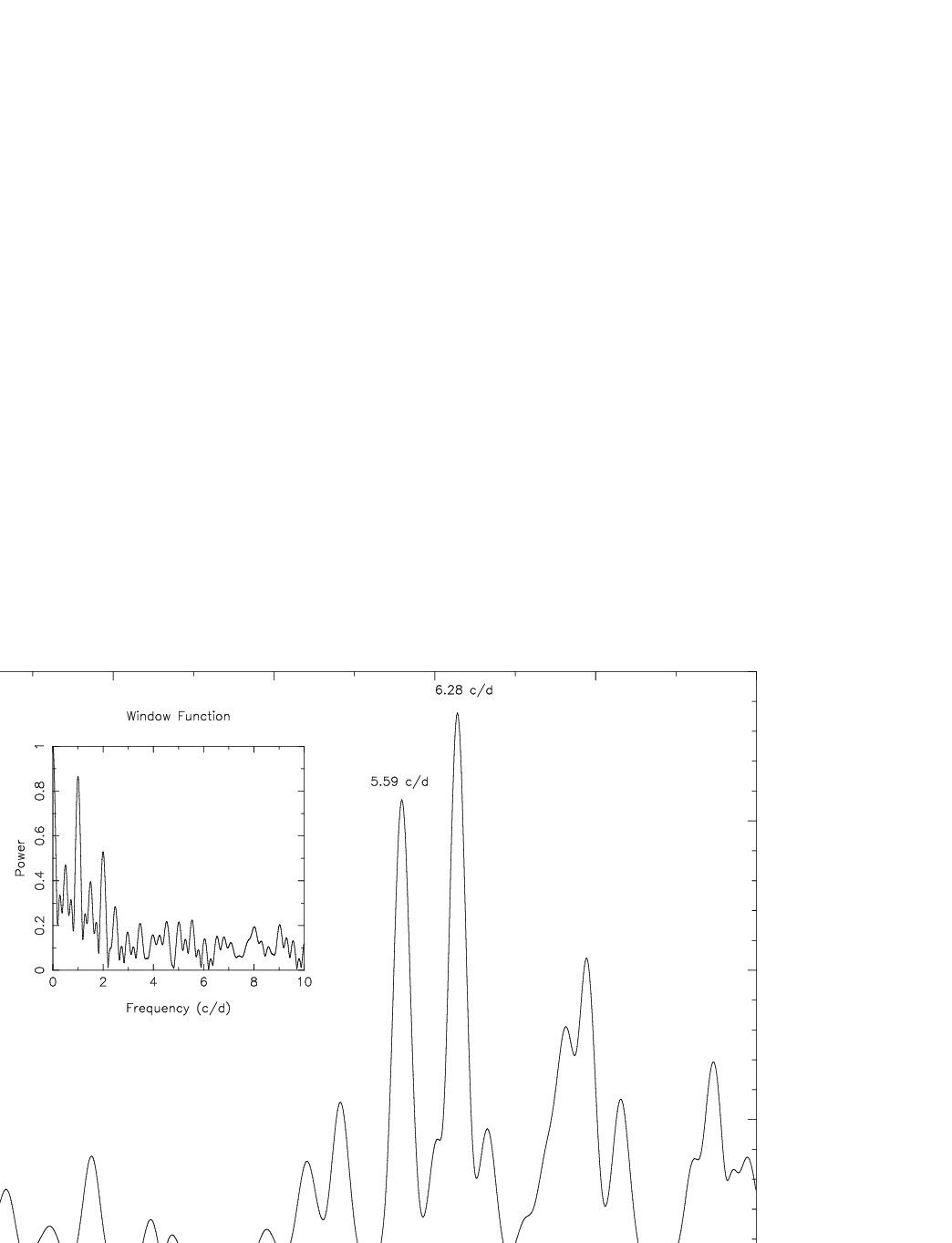}

\caption[]{\label{fig.clean} {Power spectrum of radial velocity
measurements of \bCen{}.  The upper panel shows the raw power spectrum and
the lower panel shows the results after CLEANing.  The inset shows the
power spectrum of the window function.}}

\end{figure*}

\fi

The two-component fit to the average spectra also yields equivalent widths
of $250 \pm 10$ m\AA\ and $140 \pm 10$ m\AA, respectively.  The tables by
\citeone{B+B90} list non-LTE equivalent widths for Si\,II, III, and IV as a
function of temperature, gravity, abundance, and turbulent velocity.
Considering all possible ranges for the abundance and the turbulent
velocity for the two components, and allowing for the equivalent width
uncertainties, these tables give the following ranges for $T_{\rm eff}$ and
$\log g$, respectively: 21\,000--25\,000\,K and 3.1--4.5 for the primary
and 19\,000--21\,000\,K and 3.3--4.5 for the secondary.  These temperature
and gravity estimates agree with the values derived from Str\"omgren
photometric data, assuming a single star \citeeg{Sho85}.  Our temperature
estimate for the secondary argues against \citename{S+R68}'s
(\citeyear{S+R68}) suggestion that the companion is hotter than the
primary.  Furthermore, the temperature difference between the two stars is
consistent with zero, and is therefore also consistent with the limits on
the magnitude difference found from our interferometric observations.

\subsection{Line-profile variations}
\label{Line-profile variations}

Our spectra show clear indications of line-profile variability.  While the
temporal coverage is limited (50 spectra over five consecutive nights), it
is more complete than previous studies (see below).  This encouraged us to
undertake a period search in our data.  The line profiles in individual
spectra are asymmetric, making them unsuitable for Gaussian fitting, so we
have chosen instead to calculate the first velocity moment We did this for
the stronger line using the procedure given by \citeeg{ADPW92} and the
integration limits shown in Fig.~\ref{fig.spectrum}. Analysis of the weaker
line is discussed below.

\if\preprint1
	\begin{figure*}

\hfill
\includegraphics[width=\the\hsize*\real{0.35},bb=109 67 379 351]
{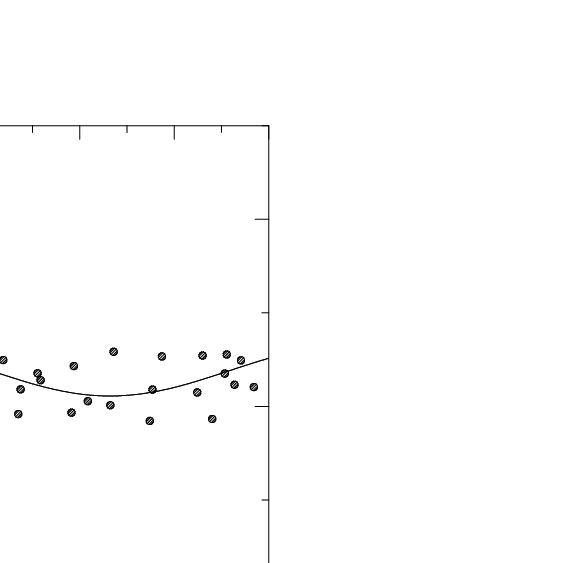}
\hfill
\includegraphics[width=\the\hsize*\real{0.35},bb=109 67 379 351]
{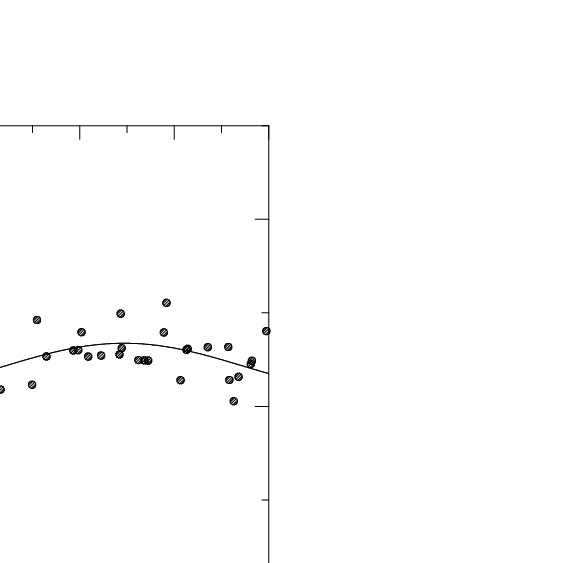}
\hfill~



\caption[]{\label{fig.rvcurves} Radial-velocity curves measured from the
stronger spectral line, phased to the frequencies $f_1=5.59$\,c/d (left)
and $f_2=6.28$\,c/d (right).  The dots represent the data, while the full
line is the fit.  We have used the same scale as \citeone{S+R68} and
\citeone{Lom75} in order to be able to compare the scatter in the diagrams
with previously suggested models.}

\end{figure*}

\fi

\if\preprint1
	\begin{figure*}

\hfill
\includegraphics[draft=false,width=\the\hsize*\real{0.45},bb=179 125 544 507]
{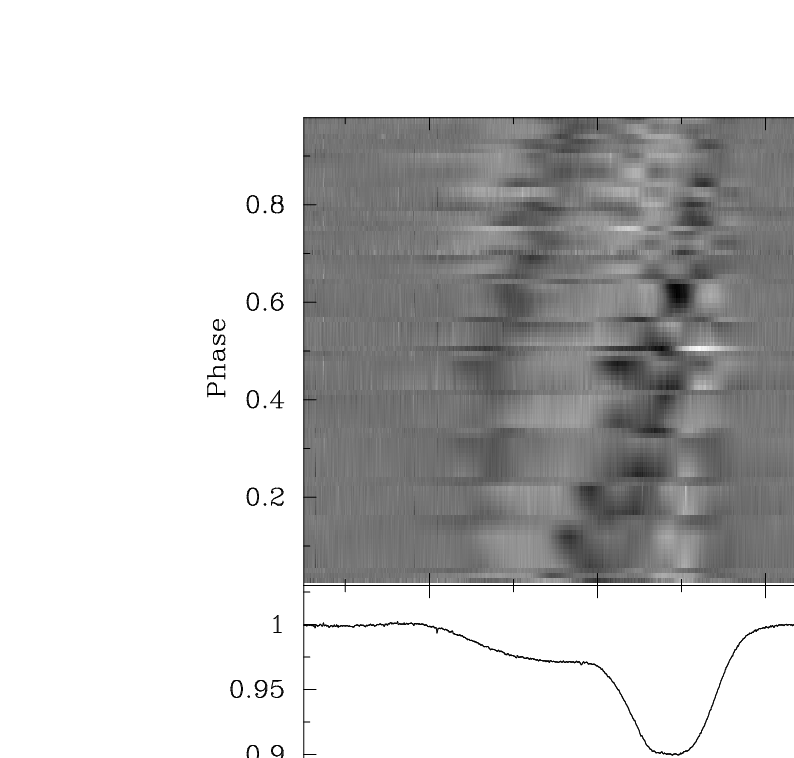}
\hfill
\includegraphics[draft=false,width=\the\hsize*\real{0.45},bb=179 125 544 507]
{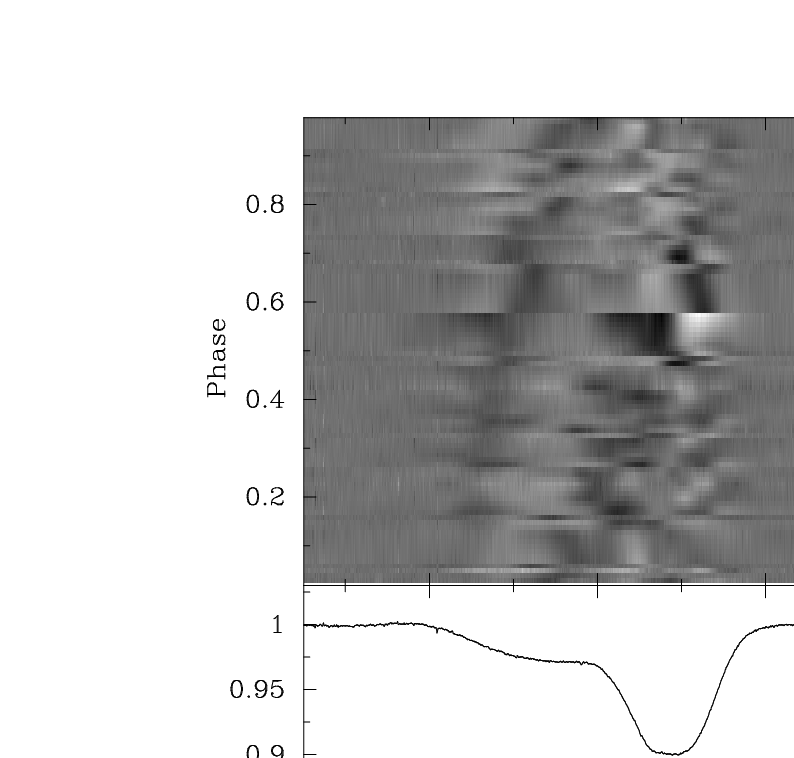}
\hfill~

\caption[]{\label{fig.greyscale} Grey-scale representations of our spectra,
phased to the frequencies $f_1=5.59$\,c/d (left) and $f_2=6.28$\,c/d
(right), with respect to the reference epoch HJD\,2447000.  The grey-scale
shows residuals with respect to the average profile (shown in the lower
panels), with black corresponding to local absorption and white indicating
local emission.}

\end{figure*}

\fi

The upper panel of Fig.~\ref{fig.clean} shows the power spectrum of the
velocity measurements.  There is an excess of power centred at 6\,c/d, with
the two strongest peaks being at 5.6 and 6.3\,c/d.  With each of these are
associated sidelobes at $\pm 1$\,c/d, which we have removed using the CLEAN
algorithm \cite{RLD87}.  The result is shown in the lower panel of
Fig.~\ref{fig.clean}.  Two peaks with an excess power compared to the other
peaks occur in the periodogram. Their significance level is similar to that
of the 6.37 c/d peak found by \citeone{Lom75}. The two peaks have
frequencies $f_1=5.59$\,c/d and $f_2=6.28$\,c/d and are about equal in
power. Taking into account the results found previously for the star (see
Section \ref{Comparison with previous studies}), we believe that these two
peaks probably represent the true frequencies of radial velocity
oscillation.

Phased radial-velocity curves for both frequencies are shown in
Fig.~\ref{fig.rvcurves}.  A model based on both frequencies accounts for
60\% of the variance present in the radial velocity of the stronger line,
implying that more modes are present at lower amplitudes. The high quality
of the spectra rules out the remaining variance being due to the noise
level of the data.  This model gives amplitudes of
$2.2\pm0.4$\,km\,s$^{-1}$ and $2.0\pm0.4$\,km\,s$^{-1}$ for $f_1$ and
$f_2$, respectively, and an average radial velocity for the primary of
$63.9\pm0.3$\,km\,s$^{-1}$.

Figure~\ref{fig.greyscale} shows grey-scale representations of our data
when folded at $f_1$ and $f_2$ (reference epoch HJD 2447000).  The figure
shows residuals with respect to the average profile.  It is clear from
these representations that $\beta\,$Cen shows line-profile variability and
that the patterns point to a complicated multi-periodic pulsation.  The
left panel of Figure ~\ref{fig.greyscale}, phased at frequency $f_1$, shows
a clear progression of line residuals from blue to red, which is a
signature of non-radial pulsations \cite{Yan}.  The progression is less
clear when the data are phased using $f_2$, showing that variations at
$f_1$ dominate.

In the grey-scale representations we also see variability in the spectrum
at the wavelength of the secondary's Si\,III line.  In an attempt to
determine whether this effect is due to contamination by
the variation of the lines of
the primary or to intrinsic variability of the secondary, we found the
power spectrum of the velocity variations of the secondary's line, using
its first moment as the velocity indicator. However, due to the greater
width of the secondary line, and blending with the primary line, the
results are uncertain. The power spectrum shows the same two frequencies
$f_1$ and $f_2$. This could be caused either by line blending leading to
contamination by the primary's variation, or, more interestingly, by the
secondary exhibiting similar pulsation modes. If this latter speculation is
true, Figure ~\ref{fig.greyscale} suggests that the variations of the
secondary may be tidally locked to those of the primary.  Finally, we
determined the average radial velocity for the secondary to be
$-106.3\pm0.5$\,km\,s$^{-1}$ at the epoch of our observations (1988 May 16-20).

\subsection{Comparison with previous studies}
\label{Comparison with previous studies}

Three previous studies of radial velocity variability in this star have
been published.  \citeone{Bre67} reported short-term variability with a
period of either 0.1317\,d (7.59\,c/d) or 0.1520\,d (6.58\,c/d), with the
ambiguity being due to 1\,c/d aliasing (only one-night runs were obtained).
\citename{Bre67}'s suggested frequencies would appear to be aliases of our
$f_1$.

{}From a set of 22 spectra taken over four nights in 1967, \citeone{S+R68}
found the most likely period to be 0.1348\,d (7.42\,c/d).  {}From a set of
39 spectra taken over two nights in 1969, \citeone{Lom75} found a period of
0.157\,d (6.37\,c/d).  He interpreted the period found by \citeone{S+R68}
as a 1\,c/d alias.  Given the short time spans (and hence, poor frequency
resolution) of the observations, the frequency found by \citeone{Lom75} is
consistent with our $f_2$.

We conclude that $\beta\,$Cen probably has at least two pulsation
frequencies.  Both frequencies found in earlier spectroscopic data appear
to be simultaneously present in the radial velocity of the star.  The beat
period amounts to 1.45 days.  The quality of the data and the limited time
spread are possible reasons why the star was found to be mono-periodic in
earlier studies.  Another explanation could be that the modes change
strength during the orbital motion due to tidal forces.  Since the orbital
parameters are not known for $\beta\,$Cen, we are not able to check if the
dates of the observations obtained by the various authors support such a
picture.  In order to derive accurate spectroscopic orbital parameters,
$\beta\,$Cen has been included in our observational programme of the search
for forced oscillations in pulsating stars (see \citebare{ADMDC97}).

\section{Hipparcos data}

$\beta\,$Cen is one of 13 known $\beta\,$Cep stars that are classified as
unsolved in the Hipparcos catalogue.  This means that the known period of
0.157\,d is not recovered in the Hipparcos photometry.  This is not
surprising, since earlier ground-based photometric surveys have also failed
to find a regular variability pattern for this star.  The only report of
photometric variability is from Str\"omgren photometry by \citeone{Bal77},
who found a 0.30-d periodicity on one night but no variability on a second
night.  Balona (priv.\ comm.) observed the star again in 1988 and 1990 and
kindly made his data available.  We did not find any clear periodicity in
these data.

\if\preprint1
	\begin{figure}

\includegraphics[width=\the\hsize,bb=50 64 456 336]{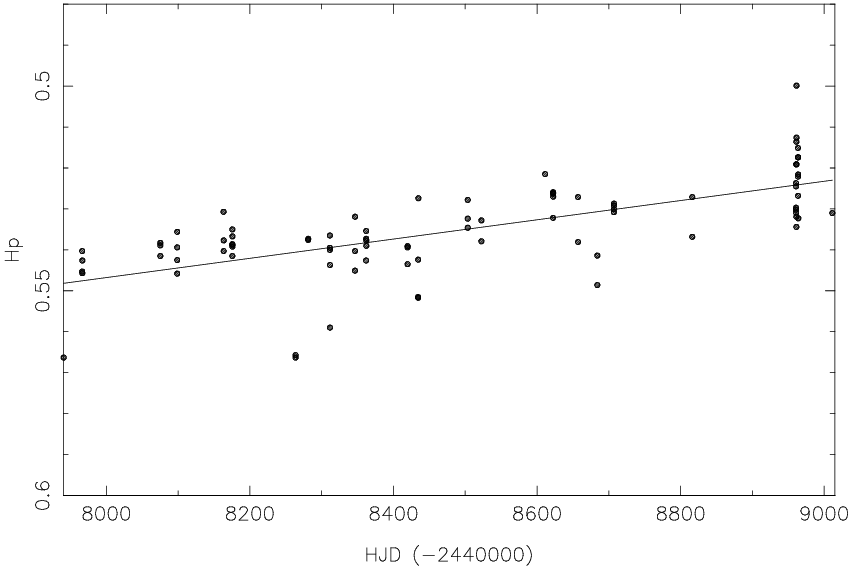}

\caption[]{\label{fig.hipparcos} The Hipparcos photometric data (dots) with
quality label $\leq\,2$ for $\beta\,$Cen.  The full line represents the
least-squares linear fit, which we have subtracted before searching for
periodicities in the data}

\end{figure}

\fi

Figure~\ref{fig.hipparcos} shows the Hipparcos photometry for \bCen{},
including only those data points having quality label $\leq\,2$ (for an
explanation of the quality labels of the Hipparcos data, see
\citeone{ESA97}). There is a clear linear trend, due to instrumental effects,
which may have impeded the pipeline analysis. We have therefore re-analysed
the Hipparcos photometry by removing the linear trend and looking for
periodicity in the residuals by means of the PDM, Scargle, and CLEAN
algorithms.  We still find no evidence of a clear periodicity, a negative
result that is expected for pulsation modes with a relatively high degree
$(\ell\geq3)$. Indeed, many of the known $\beta\,$Cep stars for which mode
identification has been performed and for which Hipparcos observations also
did not recover the periods of variation, do exhibit high degree
pulsation. Theoretical instability analyses predict the excitation of modes
with degrees up to $\ell=8$ in $\beta\,$Cep stars \citeeg{Mosk95}.

\section{Conclusion}

{}From interferometric measurements using the technique of non-redundant
masking, we have shown \bCen{} to be a close binary.  For the epoch 1995
Jan 13 we find the separation to be $15.6 \pm 2$\,mas and the position
angle to be either $352^{\circ}\pm 6^{\circ}$ or $172^{\circ}\pm
6^{\circ}$.  The parallax of \bCen{}, as measured by the Hipparcos
satellite, is $6.21 \pm 0.56$\,milliarcsec \cite{ESA97}, which implies that
the physical separation of the binary at the epoch of our MAPPIT
observations was at least 2.5\,AU.

Our spectroscopic measurements reveal \bCen{} to be double-lined.  We
detect two pulsation frequencies from radial-velocity variations of the
primary star.  Further studies of this double star are underway, using both
spectroscopy (see \citebare{ADMDC97}) and also using the Sydney University
Stellar Interferometer (J. Davis et al., in prep.).  The combined data
should allow determination of the orbital parameters and component masses
of this system.

\section*{Acknowledgements}

We thank the AAT staff for their help in setting up MAPPIT, and the night
assistant Steve Lee.  We thank John Davis for suggesting we observe \bCen{}
with MAPPIT and Luis Balona for making available his unpublished
time-series photometry.  The development of MAPPIT was supported by a grant
under the CSIRO Collaborative Program in Information Technology, and by
funds from the Australian Research Council. We thank the referee, Chris
Haniff, for comments that have improved the paper.

\if\preprint0
	\clearpage
	\renewcommand{\baselinestretch}{1}
        \begin{table*}
\begin{flushleft}
\caption[]{\label{table.obs} Journal of MAPPIT observations of \bCen}
\begin{tabular}{llllllll}

Date & $\lambda$/$\Delta\lambda$ & Mask &  Longest & Hole & 
		Detector& Observed position angles & Seeing\\
     &           & pattern & baseline$^*$ & size$^*$ & field & (North through East)& \\
(UT) & (nm)& & (m)& (cm) & (arcsec) & (degrees)& (arcsec) \\ \hline
1995/1/12 & 650/40 &~A& 3.28 & 4.7 & 3.3 x 5.8 & 87.8, $-$31.1, 29.8 & 1.1 \\
1995/1/12 & 650/40 &~B& 3.67 & 4.7 & 3.3 x 5.8 & 5.1, 60.2, 11.1, $-$57.8 & 1.2 \\
1995/1/14 & 486/10 &~B& 3.67 & 7.9 & 2.3 x 4.1 &
	 $-$1.1, $-$90.0, $-$89.4, 45.9, $-$42.8, 4.2, $-$60.3, 
				& $\sim$1 (cloud-affected)\\
          & & & & & & 74.8, $-$13.5, 33.6, 82.2, $-$3.0, $-$71.9, $-$71.3 & \\
\hline
\end{tabular}\\
$^*$As projected on primary mirror 
\end{flushleft}
\end{table*}

	\renewcommand{\baselinestretch}{2}
	\clearpage
        \begin{figure}
\includegraphics[width=\the\hsize]{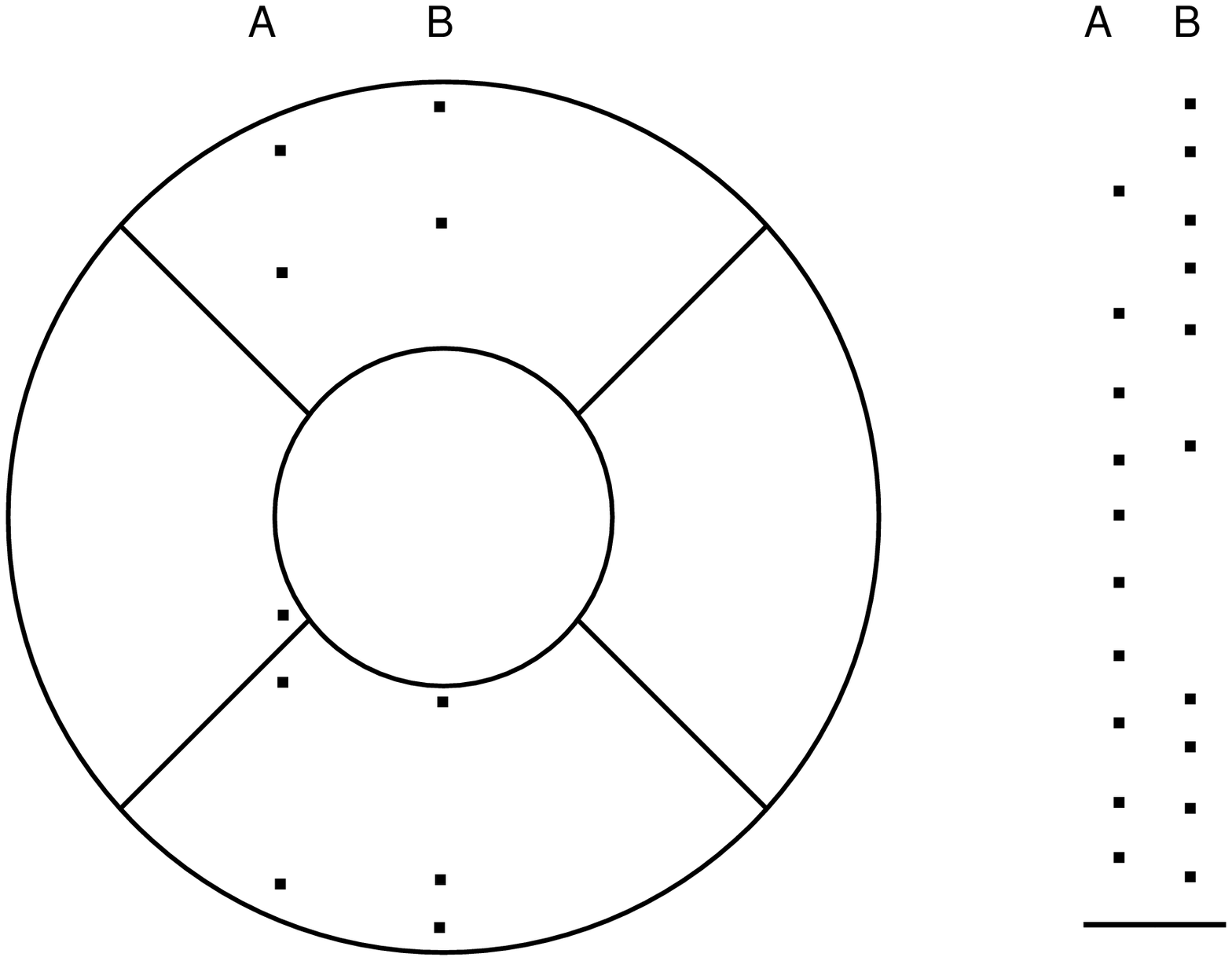}

\caption[]{\label{fig.mask} Non-redundant mask designs for MAPPIT, with the
pupil of the AAT drawn to the same scale.  The five holes in array~A have
relative positions (scaled to the longest baseline) of 0, 0.275, 0.367,
0.833 and 1.000.  The holes in array~B are at relative positions 0, 0.058,
0.275, 0.858 and 1.000.  The columns on the right show the spatial
frequency coverage of the two arrays (the line at the bottom is the
origin).  The holes as shown have sizes, as projected on the primary
mirror, of 4.7\,cm.  }

\end{figure}

        \begin{figure}

\includegraphics[width=\the\hsize,bb=89 372 556 694]{vis191.ps}

\bigskip

\includegraphics[width=\the\hsize,bb=89 372 556 694]{vis509.ps}

\caption[]{\label{fig.vis} Two examples of $V^2$ fits for MAPPIT
observations of \bCen.  The symbols show measured values and the curves are
two-parameter fits, as described in the text.  }

\end{figure}

        \begin{figure}

\includegraphics[width=\the\hsize,bb=92 372 548 694]{bfig.ps}

\caption[]{\label{fig.B} Projected angular separation of \bCen{} as a
function of position angle.  Each symbol is the result of a fit to
observations at a single position angle.  The curve shows the best-fit
theoretical model for a binary star with components of equal luminosity.  }

\end{figure}

        \begin{figure}
\includegraphics[width=\the\hsize,bb=35 232 394 451]{gaussfit.ps}

\caption[]{\label{fig.spectrum} The average profile of the
$\lambda\,4553\,$\AA\ Si\,III line of our $\beta\,$Cen data set of 1988 May
(full line).  The dashed and dotted lines show the best Gaussian fits for
the primary and secondary respectively.  The vertical lines denote the
integration limits for calculation of the first velocity moments of
the primary's line.}

\end{figure}

        \begin{figure*}

 \includegraphics[angle=-90, width=10cm, bb=77 51 535 707, clip=true]
{halfvrad.ps}

\smallskip

\includegraphics[width=10cm,bb=23 5 482 357]
{bcenclea.ps}

\caption[]{\label{fig.clean} {Power spectrum of radial velocity
measurements of \bCen{}.  The upper panel shows the raw power spectrum and
the lower panel shows the results after CLEANing.  The inset shows the
power spectrum of the window function.}}

\end{figure*}

        \begin{figure*}

\hfill
\includegraphics[width=\the\hsize*\real{0.35},bb=109 67 379 351]
{vrad5.ps}
\hfill
\includegraphics[width=\the\hsize*\real{0.35},bb=109 67 379 351]
{vrad6.ps}
\hfill~



\caption[]{\label{fig.rvcurves} Radial-velocity curves measured from the
stronger spectral line, phased to the frequencies $f_1=5.59$\,c/d (left)
and $f_2=6.28$\,c/d (right).  The dots represent the data, while the full
line is the fit.  We have used the same scale as \citeone{S+R68} and
\citeone{Lom75} in order to be able to compare the scatter in the diagrams
with previously suggested models.}

\end{figure*}

        \begin{figure*}

\hfill
\includegraphics[draft=false,width=\the\hsize*\real{0.45},bb=179 125 544 507]
{grey5.ps}
\hfill
\includegraphics[draft=false,width=\the\hsize*\real{0.45},bb=179 125 544 507]
{grey6.ps}
\hfill~

\caption[]{\label{fig.greyscale} Grey-scale representations of our spectra,
phased to the frequencies $f_1=5.59$\,c/d (left) and $f_2=6.28$\,c/d
(right), with respect to the reference epoch HJD\,2447000.  The grey-scale
shows residuals with respect to the average profile (shown in the lower
panels), with black corresponding to local absorption and white indicating
local emission.}

\end{figure*}

        \begin{figure}

\includegraphics[width=\the\hsize,bb=50 64 456 336]{hipparc.ps}

\caption[]{\label{fig.hipparcos} The Hipparcos photometric data (dots) with
quality label $\leq\,2$ for $\beta\,$Cen.  The full line represents the
least-squares linear fit, which we have subtracted before searching for
periodicities in the data}

\end{figure}

\else
	\bsp
\fi

\end{document}